\documentclass{book}    
\usepackage{meta}  
\usepackage{cite}
\usepackage{amsmath,amssymb}

\renewcommand{\Re}{\mathop{\rm Re}}
\newcommand{\ka}{\hbox{\ae}}

\begin{document}
\pagestyle{empty}
\title{Nonlinear-optical frequency-doubling meta-reflector: pulsed regime}
\maketitle

\author      {A. K. Popov}
\affiliation {Birck Nanotechnology Center, Purdue University,}
\address     {}
\city        {West Lafayette, }
\postalcode  {IN 47907,}
\country     {USA}
\email       {popov@purdue.edu}  
\misc        {https://nanohub.org/groups/nlo/popov }  
\nomakeauthor

\author      {S. A. Myslivets}
\affiliation {Institute of Physics, Siberian Branch of the Russian Academy of Sciences and Siberian Federal University}
\address     {}
\city        {Krasnoyarsk}
\postalcode  {}
\country     {Russian Federation}
\email       {sam@iph.krasn.ru}  
\misc        { }  
\nomakeauthor
\begin{authors}
{\bf A. K.~Popov}\,$^{1,*}$,
{\bf S. A.~Myslivets}\,$^{2,3}$\\
\medskip
$^1$\,Birck Nanotechnology Center, \/ Purdue University,
West Lafayette, IN 47907, USA\\
$^2$\,Institute of Physics, SB of the Russian
Academy of Sciences,
660036 Krasnoyarsk, Russian Federation\\
$^3$\,Siberian Federal University, 660041 Krasnoyarsk, Russian Federation\\
$^{*}$popov@purdue.edu; https://nanohub.org/groups/nlo/popov
\end{authors}
\begin{paper}

\begin{metaabstract}
The properties of backward-wave second harmonic meta-reflector operating in pulse regime are investigated.
We show that ratio of the income pulse length to the thickness of the  metaslab determines its important operational properties which are in contrast with those manifested at common  second harmonic generation settings.
\end{metaabstract}

\psection{Introduction}\label{in}
Light can be described as travelling electromagnetic wave (EMW) with phase velocity $\textbf{v}_{ph}$ directed  along the wave vector $\textbf{k}$  whereas its energy flux is represented by the Poynting vector $\textbf{S}$:
\begin{eqnarray}
\textbf{v}_{ph}= (\textbf{k}/k)(c/n), \quad
\mathbf{S}
=({c}/{4\pi})[\mathbf{E}\times\mathbf{H}] =({c^{2}\mathbf{k}}/{4\pi\omega\epsilon})H^{2}
=({c^{2}\mathbf{k}}/{4\pi\omega\mu})E^{2}.  \label{s}
\end{eqnarray}
Here, $c$ is speed of light,  $n$ is refractive index of the propagation medium, $\textbf{E}$ and $\textbf{H}$ are electric and magnetic components of the wave.
Refractive index is determined by   electric permittivity  $\epsilon$  and magnetic permeability $\mu$ of the  medium  at  corresponding frequency. In all naturally occurring  materials, $\epsilon>0$, $\mu>0$,  $n= \sqrt{\mu\epsilon}$ and vectors $\mathbf{S}$ and $\mathbf{k}$ are co-directed. In the case of $\epsilon~<~0$  and $\mu<0$, refractive index becomes negative, $n= - \sqrt{\mu\epsilon}$, and vectors $\mathbf{S}$ and $\mathbf{k}$ -- \emph{contra-directed}. Such waves are referred to as \emph{backward} EMW (BEMW). The possibility to produce backward light became achievable only recently owing to the advent of nanotechnology and optical metamaterials which has led to revolutionary breakthrough in the concept and in the numerous applications of the linear optics \cite{ShC}. Extraordinary  coherent nonlinear optical (NLO) processes  were predicted in \cite{KivSHG,Sl,APB,OL} for the cases of coupled  ordinary and BEMW waves. Among them are second harmonic generation (SHG)  \cite{KivSHG,Sl,APB}, optical parametric {amplification} and frequency-shifted {nonlinear reflectivity} \cite{APB,OL,Mir}.
{Metamaterials} (MM) are artificially designed and engineered materials, which can have properties unattainable in nature.
Current mainstream in fabricating negative-index MM (NIM) slabs relies on engineering of LC nanocircuits - plasmonic mesoatoms at the nanoscale with negative electromagnetic response. Extraordinary coherent nonlinear optical frequency-converting propagation processes predicted in NIMs have been experimentally realized to date in the microwave \cite{Mw} and in the multilayered plasmonic optical MM \cite{Such}.
However, a different, more general approach, to engineering MM that can support BEMW is possible \cite{Agr,AgGa}.  It is based on the fact that in a loss-free isotropic medium, energy flux $\mathbf{{S}}$ is  directed along the group velocity ${\mathbf{v}_g}$:
\begin{equation}
{\mathbf{S}}={\mathbf{v}_g}U, \/ {\mathbf{v}_g}=\rm grad_{\mathbf{k}}\omega(\mathbf{k}). \label{gr}
\end{equation}
Here, $U$ is energy density attributed to EMW. It is seen that  group velocity  is directed \emph{against} the wavevector if dispersion $\partial\omega/\partial k$ becomes negative. Basically, negative dispersion $\partial\omega/\partial k<0$ can appear even in fully dielectric materials with particular composition of its structural elements. This opens an entirely novel research and application avenue. Various particular realizations of negative dispersion were proposed so far \cite{Tr,NT,APA2012,AST2013,SSP2014}. Basically, many hyperbolic MM  and specially designed waveguides can support BW  electromagnetic modes, see, e.g., \cite{Nar,Mok,Chr}. This opens new avenues for creation of BW photonic devices with unparalleled functional properties. As regards coherent nonlinear-optical propagation processes, critically important is to provide for coexistence of coupled ordinary and backward  waves which frequencies and wave vectors satisfy to energy and momentum conservation law (phase matching). Such possibilities were described in ref.~\cite{APA2012,AST2013,SSP2014}.

This paper is to investigate unusual dependence of BW SHG in the pulse regime. We show that  conversion efficiency and properties of generated SH pulses depend on the ratio of the input FH pulse length to thickness of the NLO MM slab. Losses are included in the consideration.
\psection{Basic Equations}\label{pul}
As noted, properties of SHG will be investigated  for the cases where one of the coupled waves is ordinary and the other one is BW.  To achieve phase matching, wave vectors of the fundamental and the SH waves must be co-directed. This means that in BW setting pulse of the SH  will propagate \emph{against} the pulse of fundamental radiation.  Corresponding basic equations are as follows.
Electric and magnetic components of the waves and corresponding nonlinear polarizations at $\omega_1$ and at $\omega_2=2\omega_1$ are defined as
\begin{eqnarray}
\{{\mathcal{E}}, {\mathcal{H}}\}_j=\Re{\{{E},{H}\}_j\exp\{i(k_jz-\omega_jt)\}},
\{\mathcal{P}, \mathcal{M}\}^{NL}_j=\Re{\{P, M\}^{NL}_j\exp\{i(\widetilde{k}_jz-\omega_jt)\}},\label{pme}&&\\
\{P, M\}^{NL}_1=\chi^{(2)}_{e,m,1}\{E, H\}_2\{E, H\}_1^*,
 \{P, M\}^{NL}_2=\chi^{(2)}_{e,m,2}\{E, H\}_1^2,\, 2\chi^{(2)}_{e,m,2}=\chi^{(2)}_{e,m,1}. \label{pm2}
\end{eqnarray}
Amplitude $E_1$ of the first harmonic (FH) and of the SH, $E_2$,  are given by the equations:
\begin{eqnarray}
 s_2\frac{\partial E_2}{\partial z}+ \frac1{v_2}\frac{\partial E_2}{\partial t}= 
 - i\frac{k_2\omega_2^2}{\epsilon_2c^2}4\pi\chi^{(2)}_{e,2}E_1^2\exp{(-i\Delta kz)}-\frac{\alpha_2}2E_2,\label{eqE2}&&\\
  s_1\frac{\partial E_1}{\partial z}+ \frac1{v_1}\frac{\partial E_1}{\partial t}= 
  - i\frac{k_1\omega_1^2}{\epsilon_1c^2}8\pi\chi^{(2)*}_{e,1}E_1^*E_2\exp{(i\Delta kz)}-\frac{\alpha_1}2E_1.\label{eqE1}&&
 \end{eqnarray}
Here,  $v_i>0$ and $\alpha_{1,2}$  are  group velocities and absorption indices at the corresponding frequencies, $\chi^{(2)}_{\rm eff}=\chi^{(2)}_{e,2}$ is  effective nonlinear susceptibility, $\Delta k=k_{2}-2k_{1}$.   Parameter $s_j=1$ for ordinary, and $s_j=-1$ for backward wave.
With account for $k^2=n^2(\omega/c)^2$, $n_1=s_1\sqrt{\epsilon_1\mu_1}$, $n_2=s_2\sqrt{\epsilon_2\mu_2}$, we introduce
 amplitudes $e_{j}=\sqrt{|\epsilon_j|/k_j}E_j$,  $a_j=e_i/e_{10}$, coupling parameters $\ka=\sqrt{k_1k_2/|\epsilon_1\epsilon_2|} 4\pi\chi^{(2)}_{\rm eff}$ and $g=\ka E_{10}$, loss  and phase mismatch parameters $\widetilde \alpha_{1,2}=a_{1,2}L$ and  $\Delta \widetilde{k}=\Delta {k}l$, slub thickness $d=L/l$, position $\xi=z/l$ and time instant $\tau=t/\Delta\tau$. It is assumed that   $E_{j0}=E_j(z=0)$, $l=v_1\Delta\tau$ is the pump pulse length, $\Delta\tau$ is duration of the input fundamental pulse. Quantities  $|a_j|^2$ are proportional to the time dependent photon fluxes. Then Eqs. \eqref{eqE2} and \eqref{eqE1} are written as
\begin{eqnarray}
 s_2\frac{\partial a_2}{\partial \xi}+ \frac{v_1}{v_2}\frac{\partial a_2}{\partial \tau}=
  -igla_1^2\exp{(-i\Delta \widetilde{k}\xi)}-\frac{\widetilde{\alpha}_2}{2d}a_2, \label{eq2a}&&\\
    s_1\frac{\partial a_1}{\partial \xi}+\frac{\partial a_1}{\partial \tau}=
     -i2g^*la_1^*a_2\exp{(i\Delta \widetilde{k}\xi)}-\frac{\widetilde{\alpha}_1}{2d}a_1. \label{eq2b}&&
\end{eqnarray}

In the case of magnetic nonlinearity, $\chi^{(2)}_{\rm eff}=\chi^{(2)}_{m,2}$, equations for amplitudes  $H_j$ take the form:
\begin{eqnarray}
 s_2\frac{\partial H_2}{\partial z}+ \frac1{v_2}\frac{\partial H_2}{\partial t}=
 - i\frac{k_2\omega_2^2}{\mu_2c^2}4\pi\chi^{(2)}_{m,2}H_1^2\exp{(-i\Delta kz)}-\frac{\alpha_2}2H_2,\label{eq1a}&&\\
  s_1\frac{\partial H_1}{\partial z}+ \frac1{v_1}\frac{\partial H_1}{\partial t}=
  - i\frac{k_1\omega_1^2}{\mu_1c^2}8\pi\chi^{(2)*}_{m,1}H_1^*H_2\exp{(i\Delta kz)}-\frac{\alpha_1}2H_1.\label{eq1b}&&
 \end{eqnarray}
Then, equations for amplitudes
  $a_j=m_i/m_{10}$, where $m_{j}=\sqrt{|\mu_j|/k_j}H_j$, take the form of \eqref{eq2a} and \eqref{eq2b}, with coupling parameters given by $\ka=\sqrt{k_1k_2/|\mu_1\mu_2|} 4\pi\chi^{(2)}_{\rm eff}$ and $g=\ka  H_{10}$. In both cases, $z_0=g^{-1}$ is characteristic medium length required for significant NLO energy conversion. Only most favorable case of the exact phase matching $\Delta {k}=0$  will be considered below.
\psection{Extraordinary properties of the backward-wave second harmonic generation in the pulsed regime}
Major difference in SHG in ordinary, positive index materials (PIM), and BW (NIM) materials in the continuous wave (cw) regime is summarized and illustrated in Fig.~\ref{NIMPIM} \cite{Sl,APB,OQE}. In order to achieve phase matching, wave vectors for FH  and SH must be co-directed, which dictates co-directed energy fluxes in PIM [Fig.~\ref{NIMPIM}(a)] and contra-directed in NIM [Fig.~\ref{NIMPIM}(b)]. In a PIM,  photon flux in FH depleats whereas photon flux in SH grows across the medium so that the latter one may exceed that in FH by the exit of the NLO slab. In lossless PIM, \emph{sum} of the photons in SH and of the photon pairs in FH is conserved in any point of the NLO media.   On the contrary,   FH and SH propagate in the opposite directions in a NIM. Consequently, the \emph{difference} of the above indicated numbers is conserved  so that the number of photons $\hbar\omega_2$ at $z=L$ is always equal to zero and is less than the number of photon pairs in FH at $z=0$  [Fig.~\ref{NIMPIM}(c,d)]. The rate of the changes across the slab  depends on the field strength in the input FH beam [cf. Fig.~\ref{NIMPIM}(c) and (d)]. Figure~\ref{NIMPIM}(e) depicts dependence of the photon fluxes in the FH at the exit of the NLO slab and for SH at the exit, $z=L$,  for the PIM slab and photon flux of SH from the \emph{opposite}, $z=0$, edge of the NIM slab.
\begin{figure}[!t]
          \centering
         \subfigure[]{\includegraphics[width=.16\textwidth]{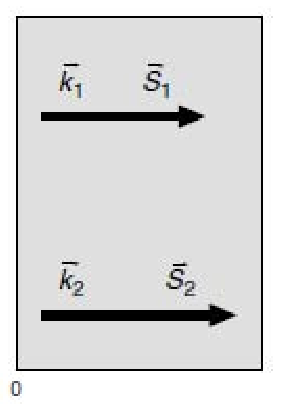}}
         \subfigure[]{\includegraphics[width=.14\textwidth]{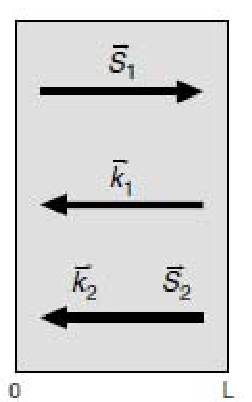}}\\
         \subfigure[]{\includegraphics[width=.3\textwidth]{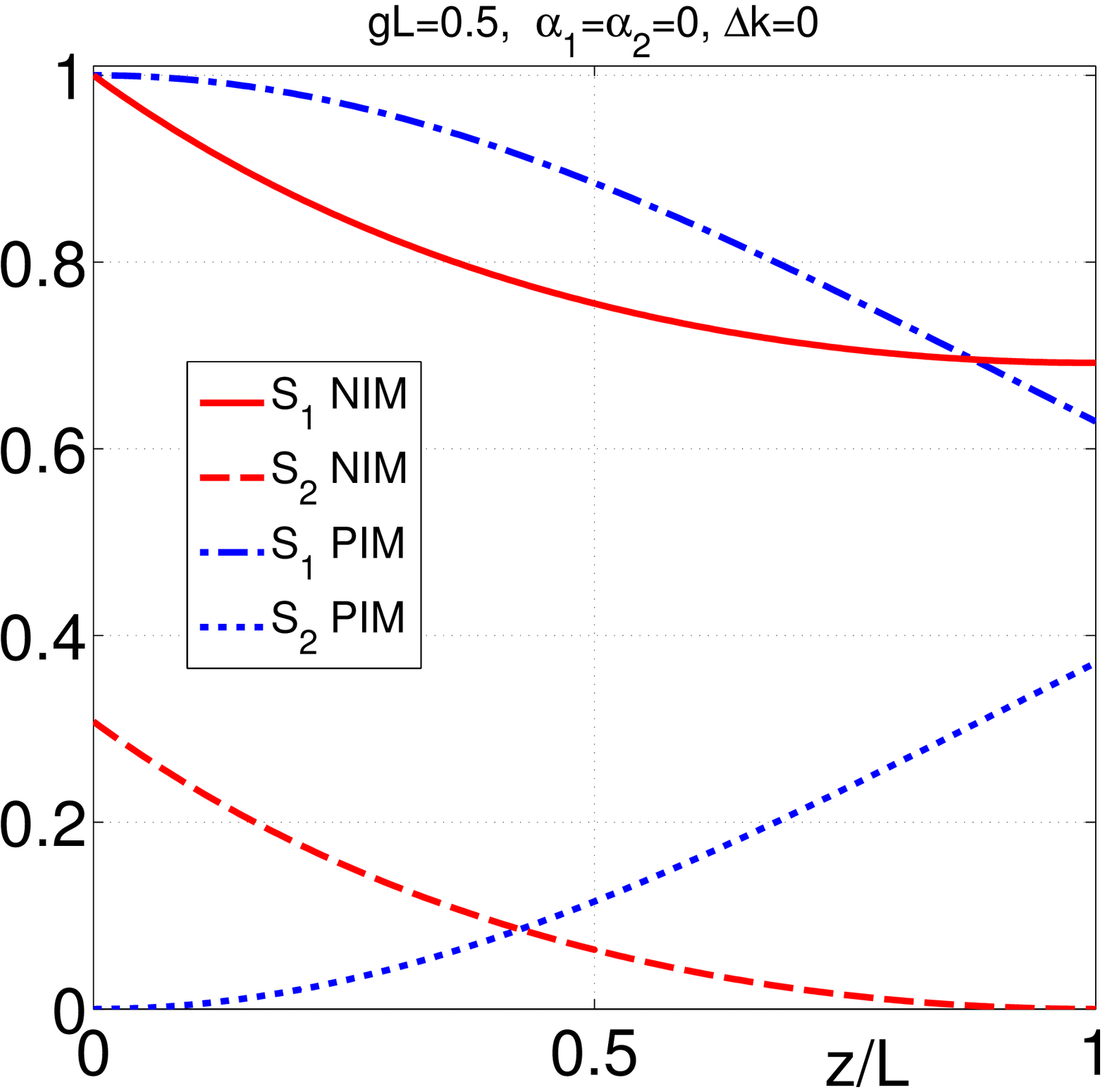}}
        \subfigure[]{\includegraphics[width=.3\textwidth]{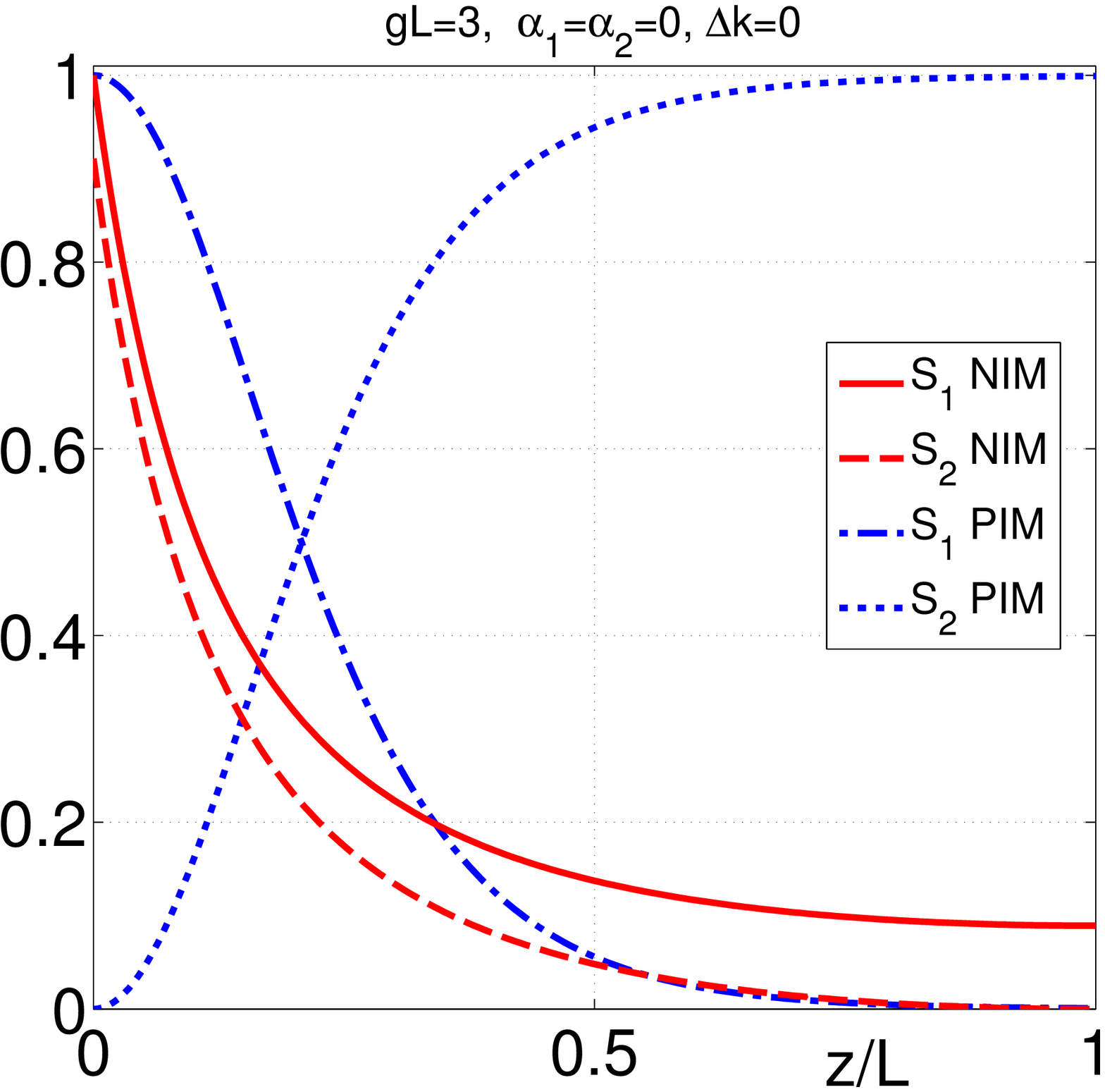}}
        \subfigure[]{\includegraphics[width=.305\textwidth]{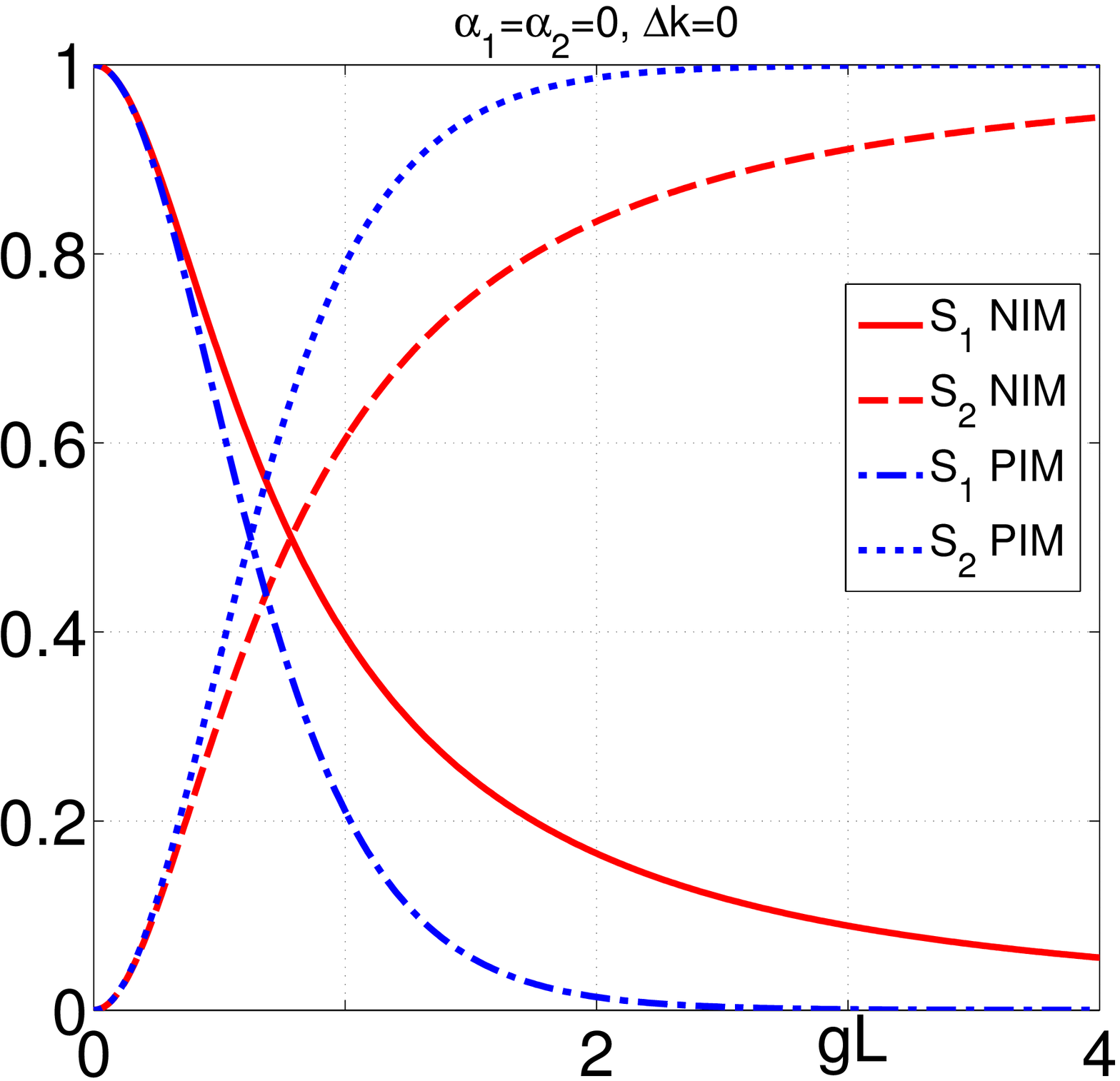}}
          \caption{Fundamental difference in the properties of continuous-wave phase matched SHG in a loss-free ordinary and  BW  materials. (a) and (b): Difference in the NLO coupling  and photon fluxes $S_{1,2}$ geometries.  (c) and (d):  Photon fluxes   reduced by the input magnitudes  across the slab. Here, the descending dash-dotted blue line depicts FH flux $S_{1}$ and  the ascending dotted blue line depicts  $S_2$  for SH, both correspond to ordinary waves in a positive-index (PIM) material. The descending solid red line depicts $S_1$  flux for \emph{backward-wave} FH, whereas  the \emph{descending} dashed red line depicts ordinary-wave  $S_2$ flux for SH, both in a frequency double-domain negative-index (NIM) slab. (c) $gL=0.5$, (d)  $gL=3$. (e): Output transmitted fundamental (the descending blue dash-dotted line) and  SH (the ascending blue dotted line) fluxes at $z=L$, both are ordinary waves in a PIM. The descending solid red line displays transmitted  BW  flux $S_1$ at $z=L$  and  the ascending dashed  red line depicts output ordinary  SH flux at $z=0$, both correspond to a NIM.}\label{NIMPIM}
\end{figure}
Obviously, the outlined properties of SHG in the case of ordinary FH and BW SH are similar. Figures~\ref{NIMPIM}(a-e) exhibit fundamental differences in the the properties of the SHG which involves backward light as compared with the SHG in ordinary crystals.

Unparalleled properties of BWSHG  in the pulsed regime stem from the fact that it occurs only inside the traveling pulse of fundamental radiation. Generation begins on the leading edge of the pulse, grows towards its trailing edge, and then exits the pulse with no further changes. Since the fundamental pulse propagates across the slab, the duration of the SH pulse is expected to be longer than that of the fundamental one. Depletion rate of the FH radiation across its pulse length and the conversion efficiency must depend on its initial maximum intensity, phase and group velocity matching. Ultimately, the overall properties of BWSHG such as the output pulse length and the photon conversion efficiency can be foreseen dependent on the ratio of the fundamental pulse and slab lengths. Investigation of the indicated  dependence is the major goal of this work.

The input pulse shape is chosen close to a rectangular form
\begin{equation}
F(\tau)=0.5\left(\tanh\frac{\tau_0+1-\tau}{\delta\tau}-\tanh\frac{\tau_0-\tau}{\delta\tau}\right),
\end{equation}
where $\delta\tau$ is the duration of the pulse front and tail, and $\tau_0$ is the shift of the front relative to $t=0$.
Parameters $\delta\tau=0.01$ and $\tau_0=0.1$ were selected for numerical simulations based on the system of partial differential equations (\ref{eq2a}) and  (\ref{eq2b}).
 The results are illustrated in Figs.~\ref{2} and \ref{3}.
A rectangular shape of the fundamental pulse, $T_1=|a_1(z)|^2/|a_{10}|^2$, is shown at $z=0$ when its leading front enters the medium and the results of numerical simulations  for the output fundamental pulse when its tail reaches the slab boundary at $z=L$. Shape and  conversion efficiency of the output  SH pulse,  $\eta_2=|a_2(z)|^2/|a_{10}|^2$, traveling against the z-axis is shown when its tail passes the slab's edge at $z=0$. Here, phase and group velocities of the fundamental and SH pulses assumed equal.
Figure \ref{2}(a) corresponds to the fundamental pulse of ten times and Fig.~\ref{2}(b) of two times longer than the slab thickness. Losses are neglected. It is seen that pulse shapes of SH are significantly different in (a) and (b) cases. It is because the transient periods when only part of the fundamental pulse is inside the slab is  longer relative its duration in the second case.
Since the metaslab is lossless, the outlined properties satisfy to the conservation law: the number of annihilated pair of photons of in the  FH ($S_{10}-S_{1L})/2$ per pulse is equal to the number of output SH photons $S_{20}$. Here, $S_{1,2 Z}$ are numbers of photons integrated over the corresponding pulse duration and  normalized by the corresponding integrated input numbers. Note, that maximum possible photon conversion efficiency for SH is 0.5. Figures \ref{2}(c,d) show effect of losses which roughly correspond to one of the model of the metaslab made of free-standing carbon nanotubes \cite{APA2012,AST2013,SSP2014}.
\begin{figure}[!t]
          \centering
         \subfigure[]{\includegraphics[width=.4\textwidth]{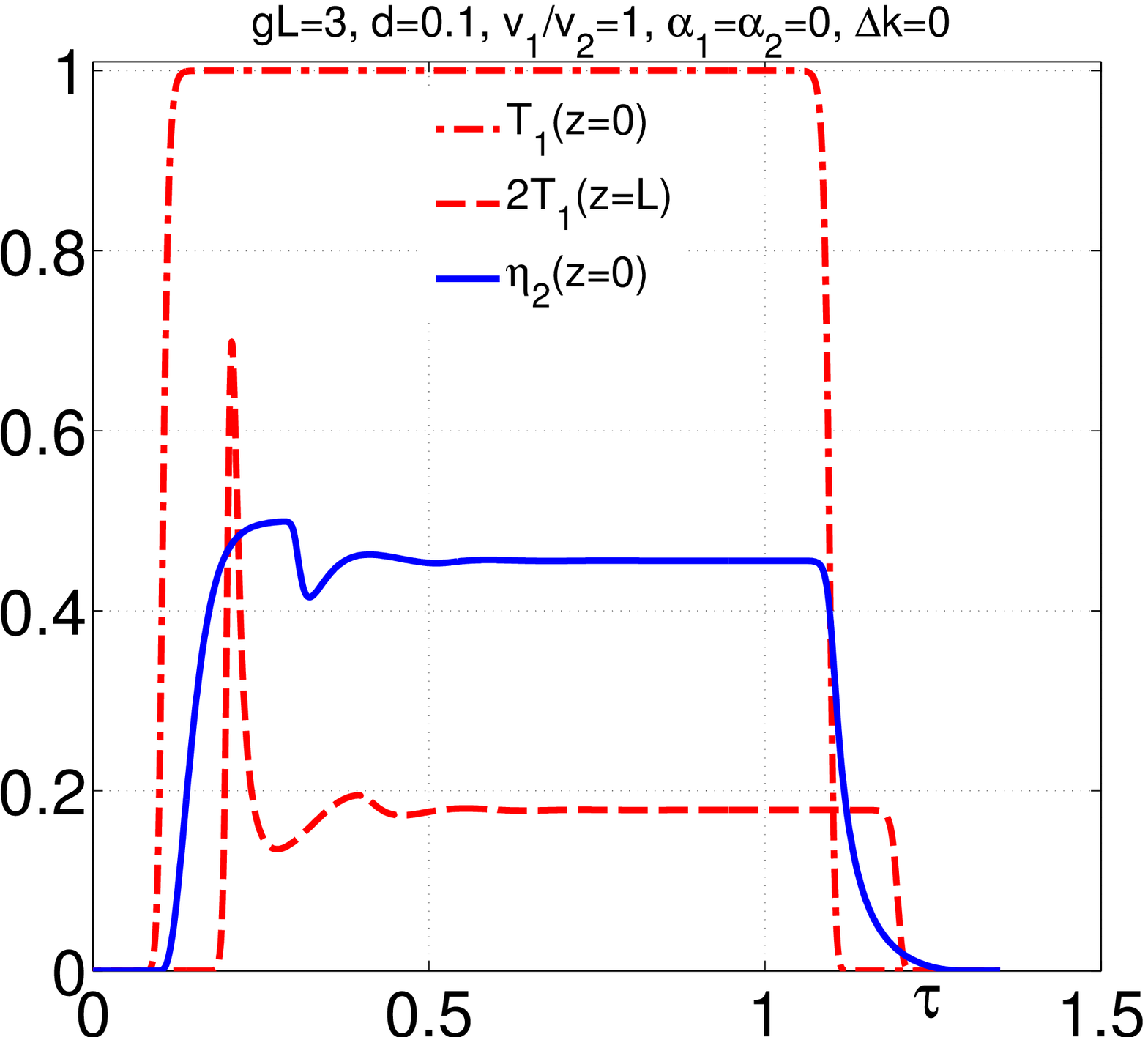}}
         \subfigure[]{\includegraphics[width=.4\textwidth]{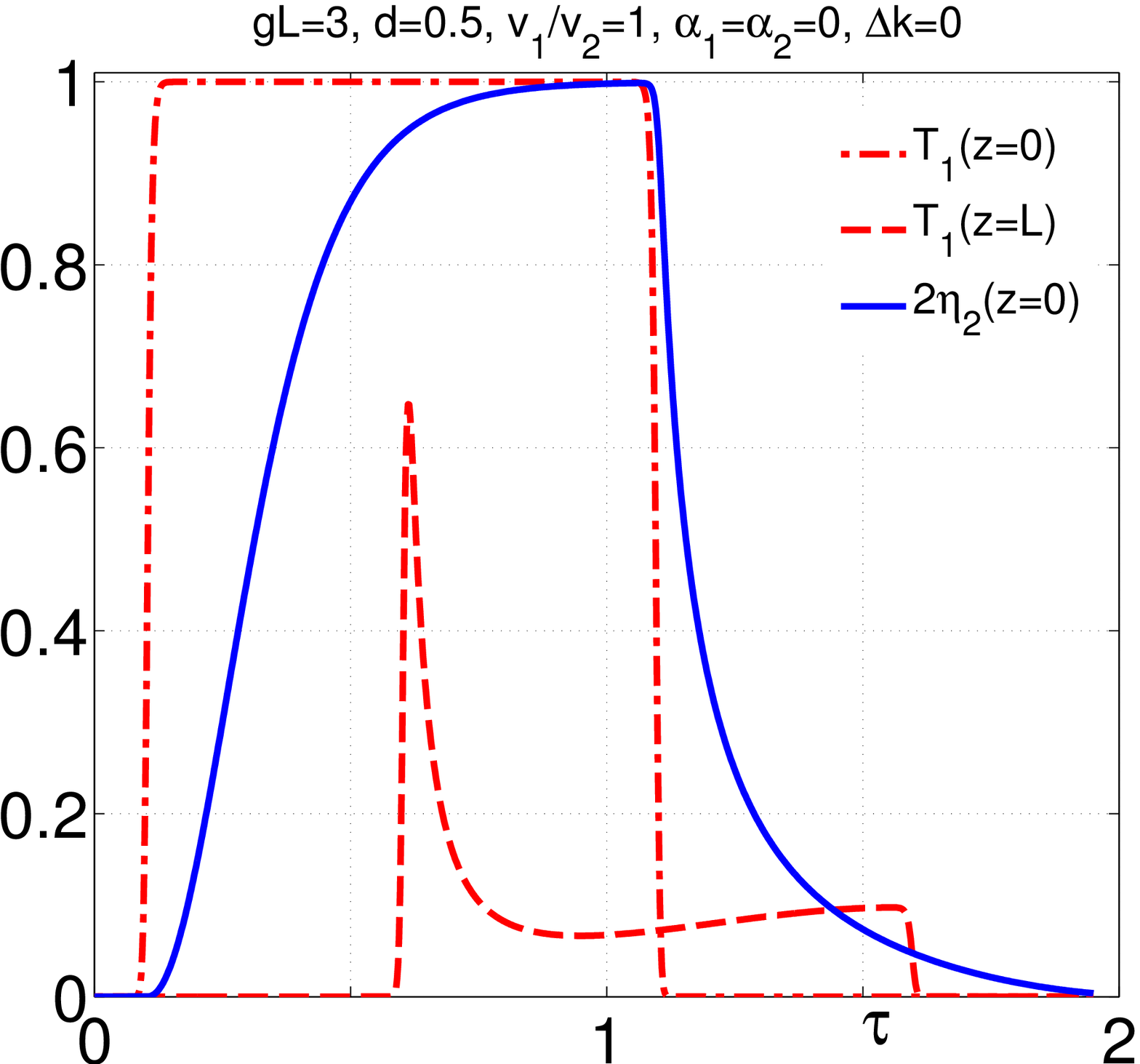}}\\
          \subfigure[]{\includegraphics[width=.4\textwidth]{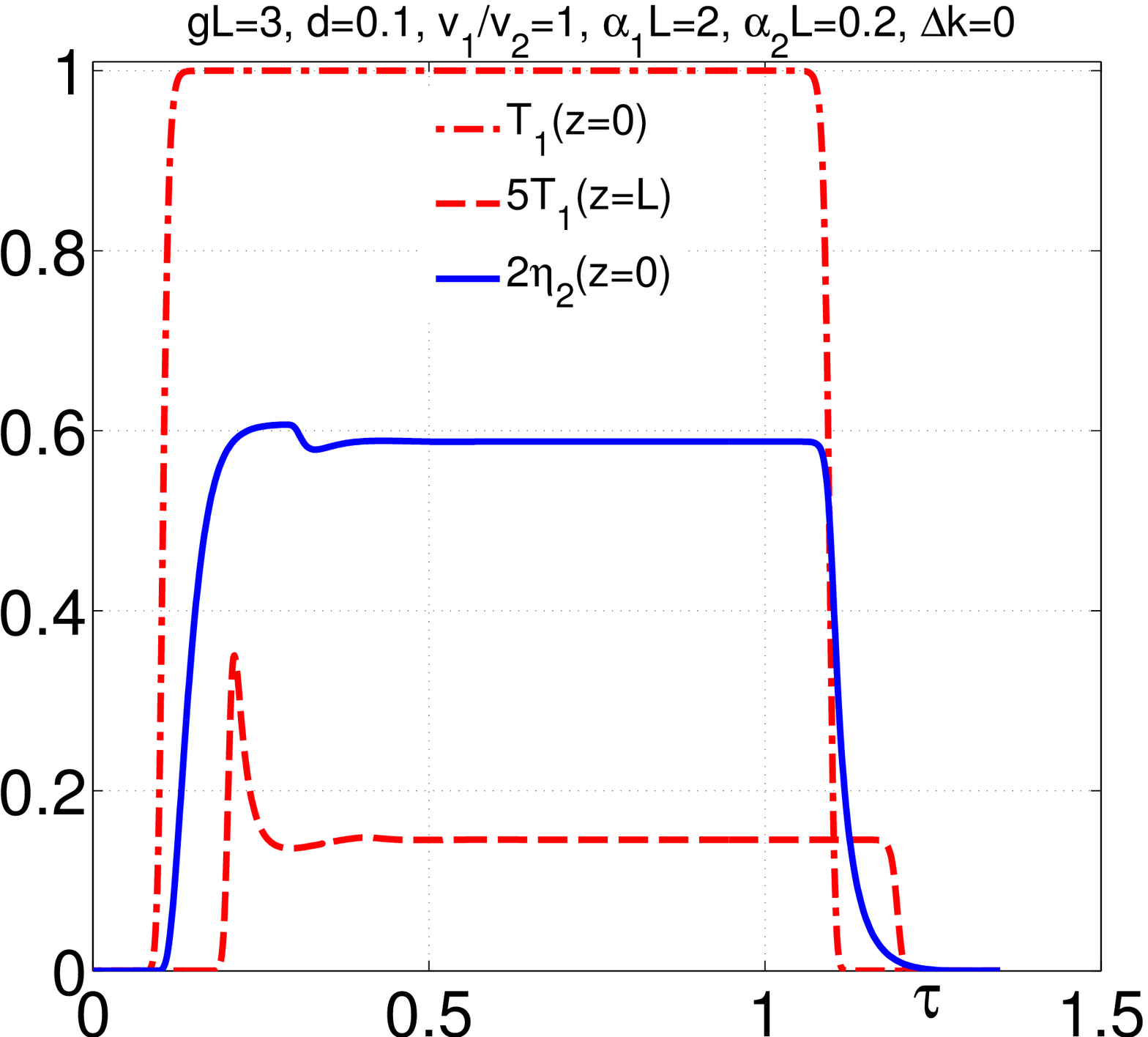}}
         \subfigure[]{\includegraphics[width=.4\textwidth]{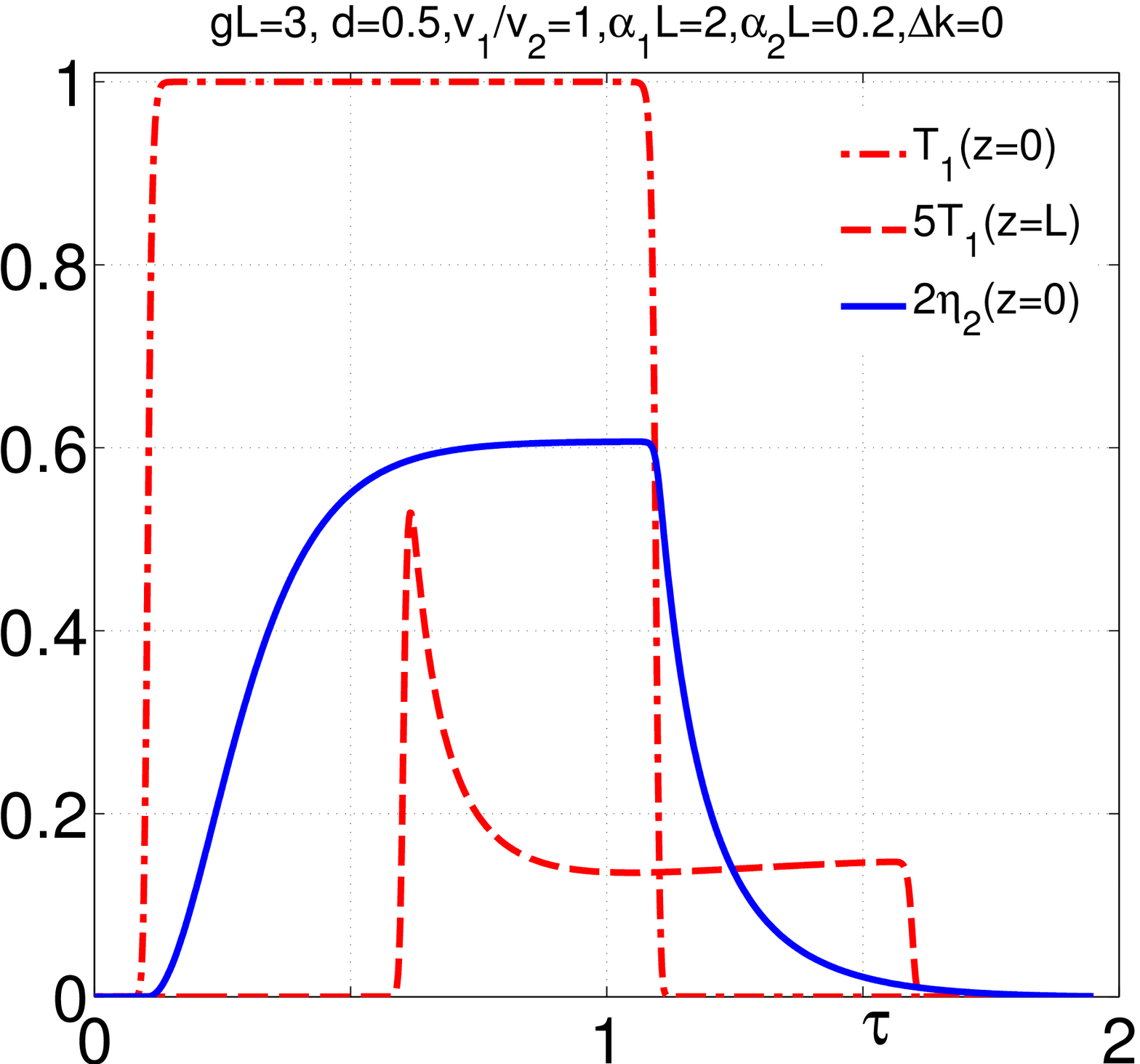}}
          \caption{Backward-wave second harmonic generation in the pulse regime.  The red dash-dotted plot depicts input (at $z=0$) pulse of FH and the red dashed plot depicts transmitted pulse at $z=L$.  The blue solid plot depicts output contra-propagating pulse at doubled frequency at $z=0$.  $\tau=t/\Delta\tau$, $gL=3$, $d=L/l$. (a, b): losses are neglected. (c, d):  $\alpha_1 L=2$, $\alpha_2 L=0.2$. (a,c):  $d=0.1$. (b,d):  $d=0.5$.
          $S_1(0)=0.9900$.
          (a) $S_1(L)=0.0934$, $S_2(0)=0.4485$.
          (b) $S_1(L)=0.1117$, $S_2(0)=0.4391$.
          (c) $S_1(L)=0.03$, $S_2(0)=0.2865$.
          (d) $S_1(L)=0.0343$, $S_2(0)=0.2695$.
          }\label{2}
\end{figure}

\begin{figure}[!h]
          \centering
         \subfigure[]{\includegraphics[width=.4\textwidth]{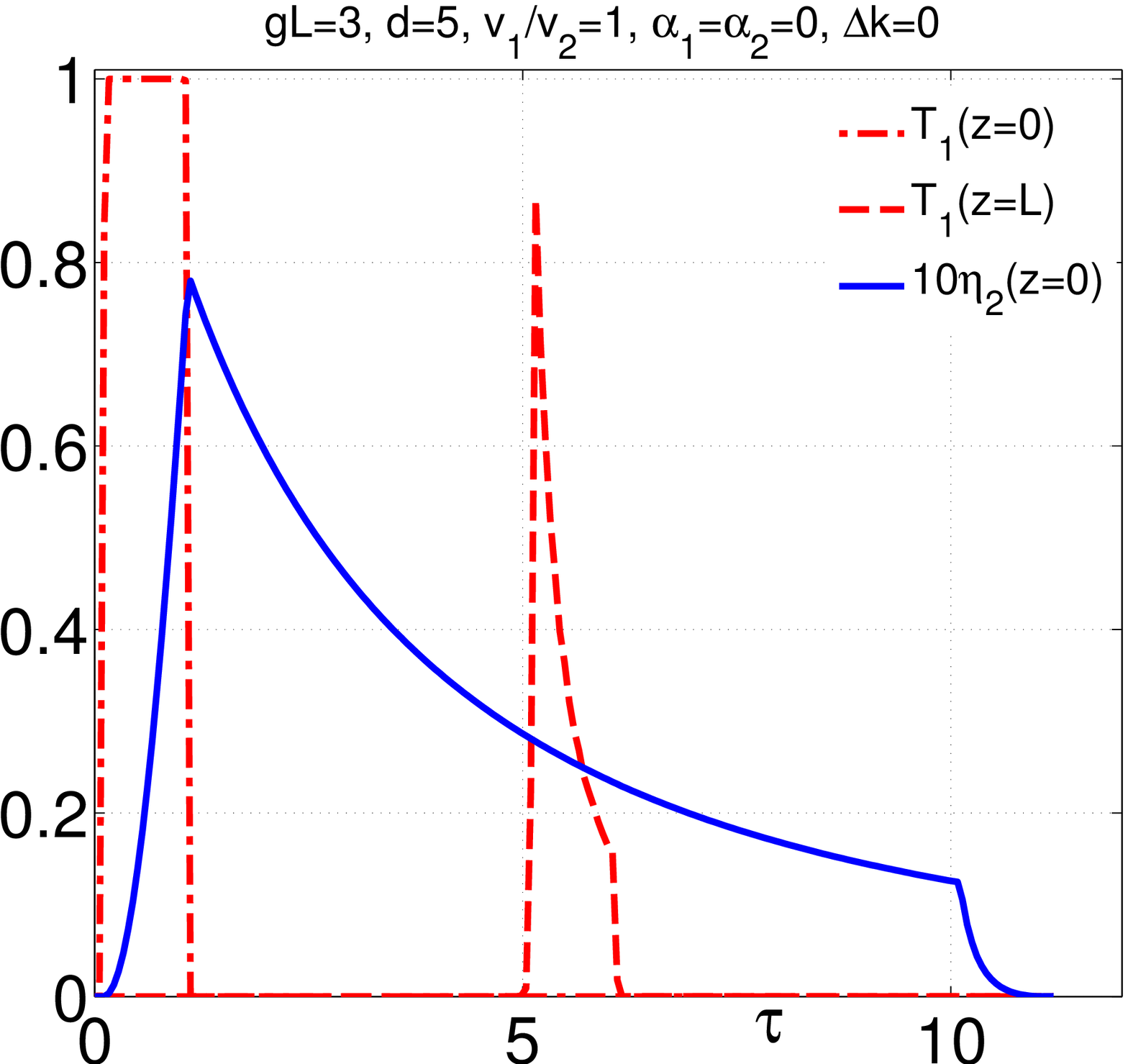}}
         \subfigure[]{\includegraphics[width=.4\textwidth]{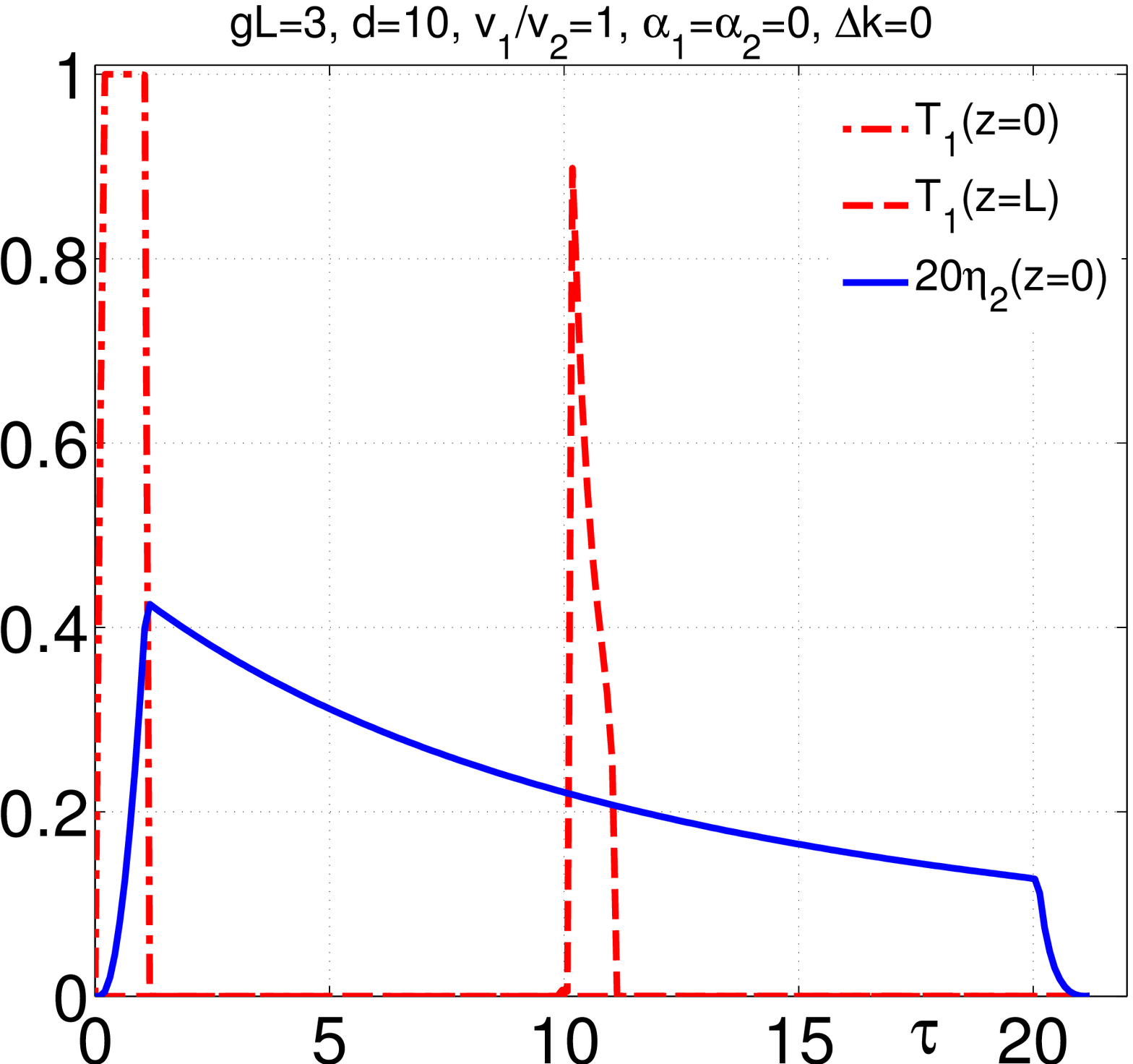}}\\
          \subfigure[]{\includegraphics[width=.4\textwidth]{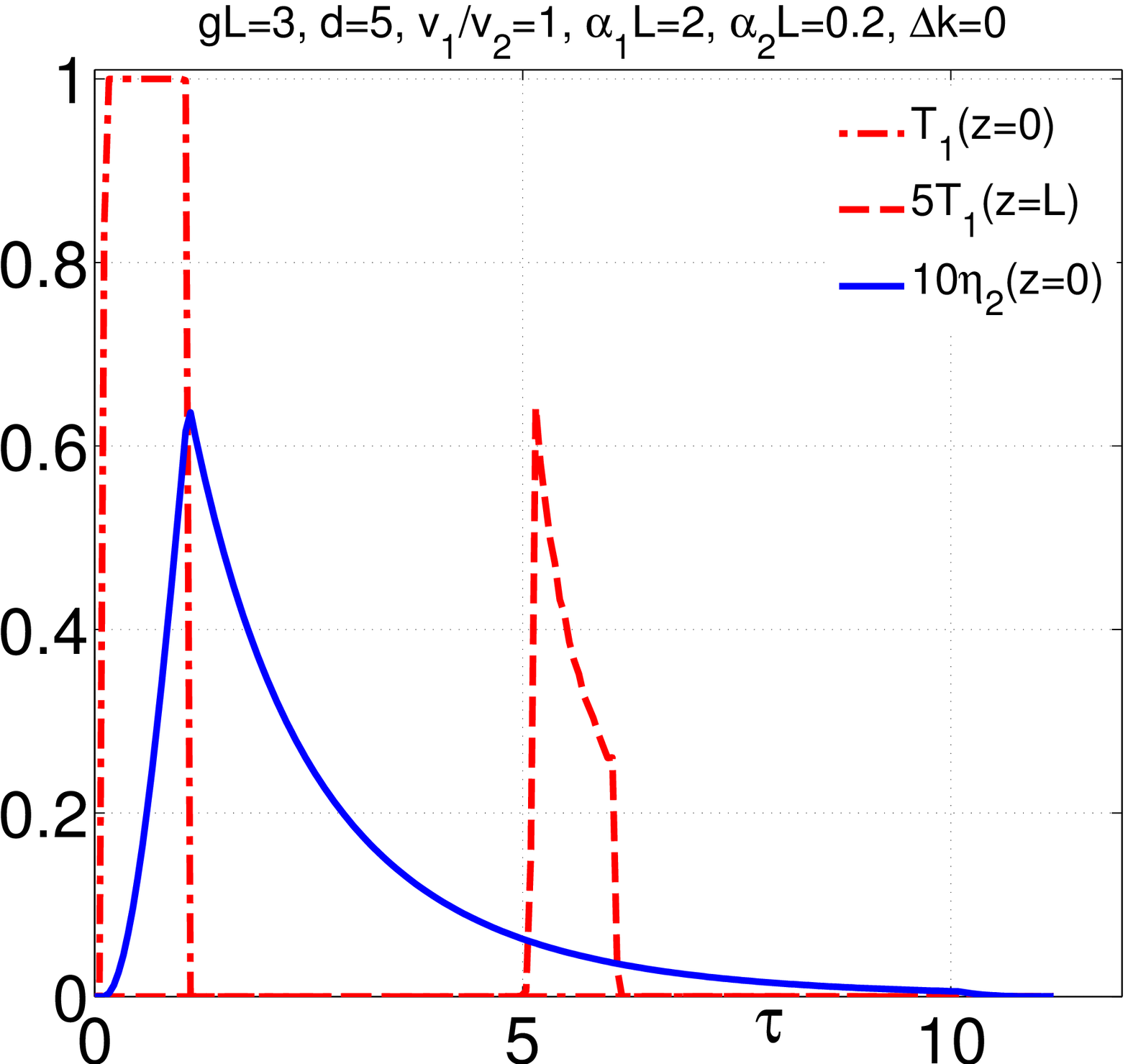}}
         \subfigure[]{\includegraphics[width=.4\textwidth]{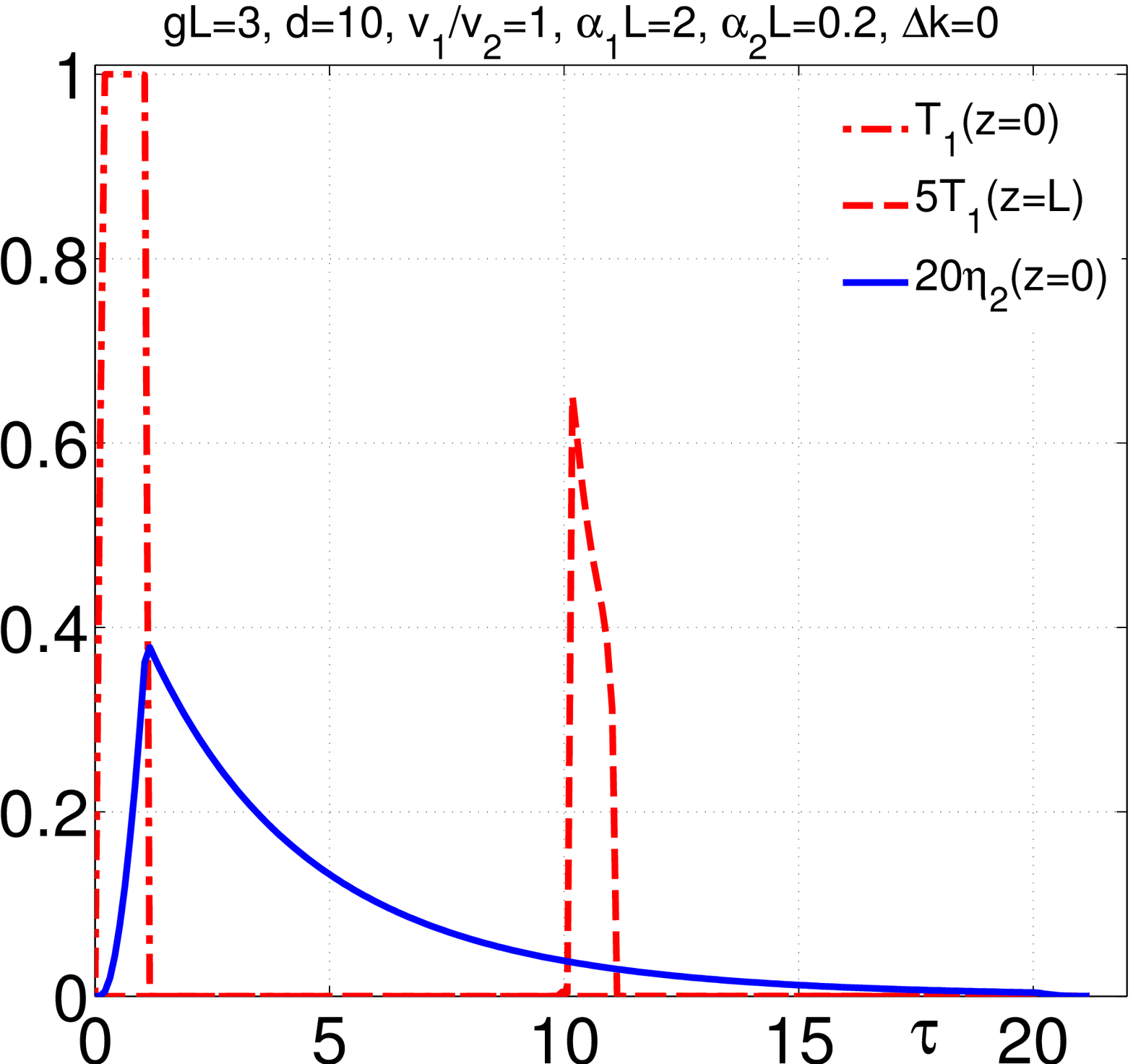}}
          \caption{ Input, transmitted and reflected pulses. (a,c):  $d=5$, (b,d):  $d=10$.
          $S_1(0)=0.9900$.
          (a) $S_1(L)=0.3607$, $S_2(0)=0.3111$.
          (b) $S_1(L)=0.5116$, $S_2(0)=0.2403$.
          (c) $S_1(L)=0.0734$, $S_2(0)=0.1273$.
          (d) $S_1(L)=0.0923$, $S_2(0)=0.0787$.
          Notations and other parameters are the same as in Fig.~\ref{2}.
          }\label{3}
\end{figure}
Figures \ref{3} (a,b) illustrate the effect of further pulse shortening. Here, Fig.~\ref{3} (a) displays input pulse five times and Fig.~\ref{3} (b) -- ten times shorter than the metaslab thickness. From comparing figures~\ref{2} and \ref{3} one concludes that difference in the pulse lengths grows so that SH pulse becomes much longer compared to the FH pulse. Counter-intuitively, quantum conversion efficiency per pulse decreases with decrease of the input pulse duration even though its maximum field strength remains unchanged and losses change SH pulse shape.
\psection{Conclusions}
Properties of frequency doubling nonlinear-optical metamirror operating in the pulsed regime are investigated. It is proposed to be engineered based on the metamaterial which supports backward light waves at one of the frequency and ordinary  wave at another frequency. It is supposed that phase velocities can be adjusted equal and  are co-directed, whereas energy fluxes of the incident and generated second harmonic light are contra-directed.  References are given to the work that prove such a possibility. Physical principles underlying fundamental difference in the properties of second harmonic generation in the proposed and standard settings are discussed. A set of partial differential equations which describe such a reflector with the account for losses are written and solved numerically. It is shown that unlike second harmonic generation in standard settings, contra-propagating generated second harmonic pulse may become much longer then the incident fundamental one and the difference  grows with decrease of the input pulse length as compared with thickness of the metaslab. The revealed properties may manifest themselves beyond the optical wavelength range.
\ack
This material is based upon work supported in part by the U. S. Army Research Laboratory and the U. S. Army Research Office under grant number W911NF-14-1-0619, by the National Science Foundation under grant number ECCS-1346547 and by the Russian Foundation for Basic Research under grant RFBR 15-02-03959A. We thank I. S. Nefedov, A. E. Boltasseva  and V. M. Shalaev for inspiring inputs.

\end{paper}
\end{document}